\documentclass[pdftex]{article}
\usepackage{icrctc07}

\title{Future imaging atmospheric telescopes: performance of possible array 
configurations for gamma photons in the GeV-TeV range}
\shorttitle{Future imaging atmospheric telescopes}
\authors{S. Sajjad$^{1}$, A. Falvard$^{1}$, G. Vasileiadis$^{1}$.}
\shortauthors{S. Sajjad and et al}
\afiliations{$^1$LPTA, Universit\'{e} Montpellier 2, CNRS/IN2P3, Montpellier, France}
\email{Saeeda.Sajjad@lpta.in2p3.fr}

\abstract{The future of ground based gamma ray astronomy lies in large arrays of Imaging Atmospheric Cherenkov Telescopes (IACT) with better capabilities: lower energy threshold, higher sensitivity, better resolution and background rejection. Currently, designs for the next generation of IACT arrays are being explored by various groups. We have studied possible configurations with a large number of telescopes of various sizes. Here, we present the precision of source, shower core and energy reconstruction for gamma rays in the GeV-TeV range for different altitudes of observation. These results were obtained through tools that we have developed in order to simulate any type of IACT configuration and evaluate its performance.}

\begin{document}
\maketitle
%Begin the section.

\section{Introduction}

 The future Imaging Atmospheric Cherenkov Telescopes (IACT) will be expected to discover new sources, enable more precise observations of known sources as well as contribute towards answering questions in adjacent fields like cosmology and particle physics. This will require large arrays of telescopes with lower energy threshold, higher sensitivity, better resolution and background rejection. Here we present the performance of two possible array configurations for $\gamma$ photons in the GeV-TeV range at two different altitudes. This study was carried out through tools that we have developed and whose description is given in the next couple of sections.

\section{Simulation tools for the study of telescope designs}
IACT systems have a large number of parameters that can be optimised to improve detection capabilites, such as the number of telescopes, their position, size and field of view and the altitude of observation. 
 In order to study IACT systems, we have developed a tool capable of simulating any type of telescope configuration and evaluating its performance.
The atmospheric showers are simulated through the CORSIKA\footnote{CORSIKA version 6.020 has been used here.} package \cite{corsika}. The reflection of the Cherenkov photons from the shower by a parabolic mirror and their impact on the telescope camera are then carried out by our IACT simulation tool.
 The program allows complete freedom in the choice of telescope diameter, focal length, camera size, photomultiplier size, telescope position and orientation and altitude of observation. Up to 100 telescopes with variable individual characteristics can be simulated at the same time so as to enable the study of very large arrays.

\section{Shower parameter reconstruction methods}

The images obtained from the simulation tools can be used to reconstruct the source position, the shower's core position on the ground (herafter simply called shower core or core) and its energy.

\begin{figure}[h!]
\begin{center}
\includegraphics [width=0.24\textwidth]{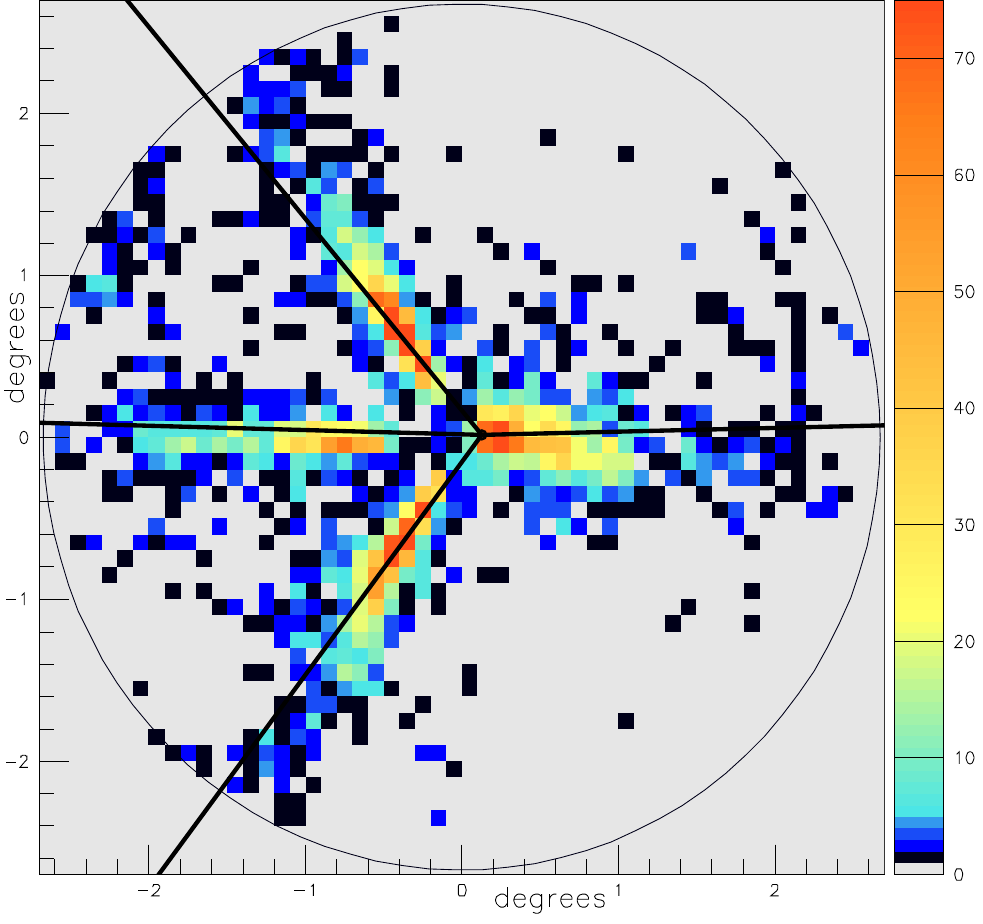}
\includegraphics [width=0.16\textwidth]{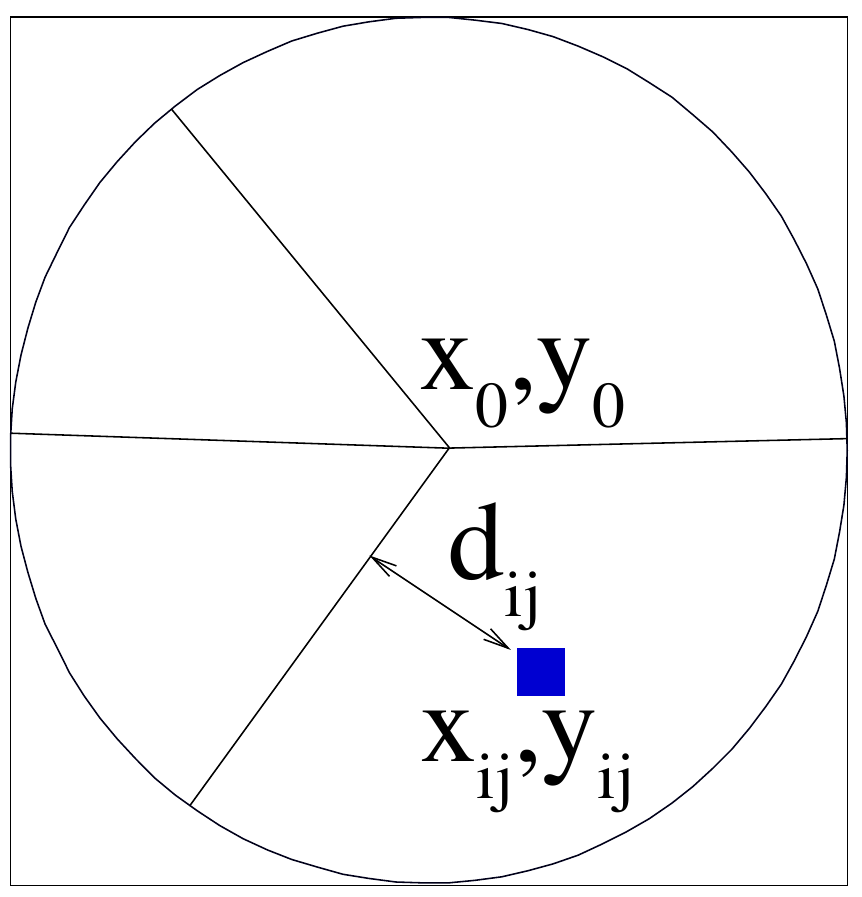}
\end{center}
\caption{The superposed images of a 500~GeV shower obtained by four telescopes. The reconstructed axis of each image is shown in black.}\label{fig1}
\end{figure}

\textbf{Source reconstruction - }
%Both the source and core reconstruction methods make use of the simultaneous information available from several telescopes observations of the same shower.
 In the camera frame of reference, each telescope gives a shower image whose axis contains the position of the source and if the images from several telescopes are superposed (see fig. \ref{fig1}), then this source position corresponds to the intersection of the axes of all images.
 
The source position can then be reconstructed by maximising the likelihood function:
\begin{equation}\label{equ1}
\mathrm{ln}(L_{all})=-\sum_{j=1}^{N_{tel}}\sum_{i=1}^{N_{pix}}\frac{N_{ij}t_{ij}^2}{2\sigma_t^2},
\end{equation}
with $d_{ij}=\frac{\left|(y_{cj}-y_o)(x_{ij}-x_o)-(y_{ij}-y_o)(x_{cj}-x_o)\right|}{\sqrt{(x_c-x_o)^2+(y_c-y_o)^2}}$. Here $N_{ij}$ is the content of the $i^{th}$ pixel from the $j^{th}$ telescope and $d_{ij}$ is its distance from the image axis. $(x_{ij}$ and $y_{ij})$ and $(x_{cj}, y_{cj})$ are the coordinates of the pixel and the centroid of the image, respectively. 
The following assumptions are made:
(1) Each image axis is a straight line passing through a point ($x_o,y_o$) common to all axes which gives the position of the source image in the camera frame of reference and whose coordinates are free parameters. 
(2) The distance of the pixels in an image from the corresponding axis (in other words the transvere image profile) is assumed to follow a Gaussian probability density function. $\sigma_t$ is the average standard deviation obtained from the Gaussian fit of transverse image profiles. (3) Each axis is made to pass through the centroid of the corresponding image.
The likelihood function is then maximised for $x_o$ and $y_o$ yields the position of the source and reconstructed image axes as shown in fig. \ref{fig1}.

\textbf{Shower core reconstruction -}
In the ground frame of reference, the point of intersection of the axes from all shower images corresponds to the core position\footnote{ Strictly true only when the source is at the zenith.}. The core position can then be reconstructed through a likelihood minimisation similar to the one performed for source reconstruction.

\textbf{Energy reconstruction -}
The energy reconstruction makes use of the linear relationship between the average number of photo-electrons obtained on a given telescope at a given distance from the shower core. Tables containing the average number of photo-electrons for a wide range of distances and shower energies have been produced and are used to deduce the energy from the photo-electron content of telescope images.

\section{Array design}

\subsection{Energy range and array parameters}

The physics goals and issues concerning $\gamma$-ray observations depend on the energy domain being considered\footnote{A discussion on the different energy domains and the scientific objectives can for instance be found in \cite{FAharonian}.}. The performance of current-day IACT indicates the following two energy domains\footnote{We have restricted this study to energies below a few tens of TeV. Beyond this limit, the flux from the sources become extremely low and very large surfaces of detection are required.}.

\textbf{High energy domain: 300~GeV - 10~TeV - } Present-day telescopes have shown that this is the domain where IACT best operate and that good angular and energy resolution can be achieved using medium sized (10-15~m diameter) telescopes. The main objective in this domain is to further increase the sensitivity of the future telescopes 
. This can be achieved by spreading a large number of telescopes over a vast area.

\textbf{Low energy domain: $<$ 30~GeV - } As the shower size decreases with energy, the images have fewer photo-electrons and are more subject to fluctuations. This makes it harder to reconstruct shower parameters and separate $\gamma$-photons from hadrons.
The main requirement in this domain is therefore the collection of a maximum amount of light from showers through the use of large sized telescopes.
At the same time, since $\gamma$-ray fluxes tend to increase at lower energies, meaningful results can be obtained with only a few telescopes.

We have chosen to work with a large number of 12.5~m diameter telescopes aimed at the high energy domain and 4 to 5 telescopes with large diameters of 30~m for the low energy domain. All telescopes have a moderate field of view i. e. 5.4$^\circ$. The square shaped pixels on the camera have a side of 0.1$^\circ$. The study is carried out at two altitudes above sea level~: 1800~m and 3600~m.

\subsection{Optimum telescope separation}
Once choices concerning telescope size and number have been made as a function of the energy domain being targeted, the optimum inter-telescope distance can be determined. In order to do so, we have chosen to study the $\gamma$-ray response of a square unit of four telescopes by uniformly generating vertical $\gamma$-rays over a large surface area. A simple trigger requiring that at least two telescopes have images with at least 50~photo-electrons is applied and shower parameters are reconstructed for the passing events.

This is carried out at 300~GeV with 12.5~m telescopes and 50~GeV with 30~m telescopes. These energies correspond to the lower ends of the high and low energy domain respectively\footnote{One notes that while the low energy domain does not have a fixed lower limit, the combination of the small size of electromagnetic showers and the higher cosmic ray background levels for lower energies seem to indicate that gamma observation below 50~GeV will be highly problematic if not impossible. We have therefore chosen to carry out the inter-telescope distance optimisation at a relatively 'safe' energy, i. e. 50~GeV.}. 

\begin{figure}
\begin{center}
\includegraphics [width=0.45\textwidth]{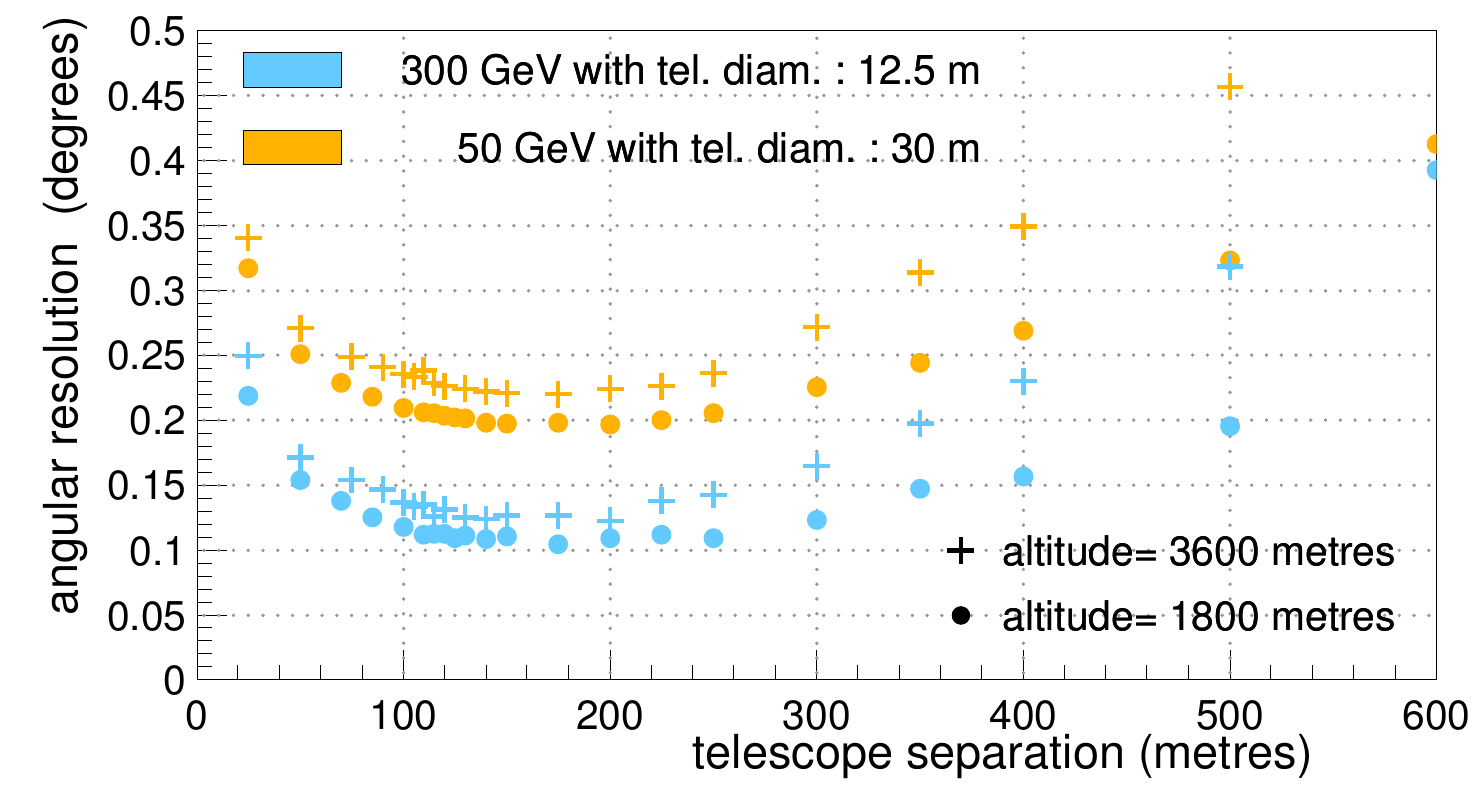}
\includegraphics [width=0.45\textwidth]{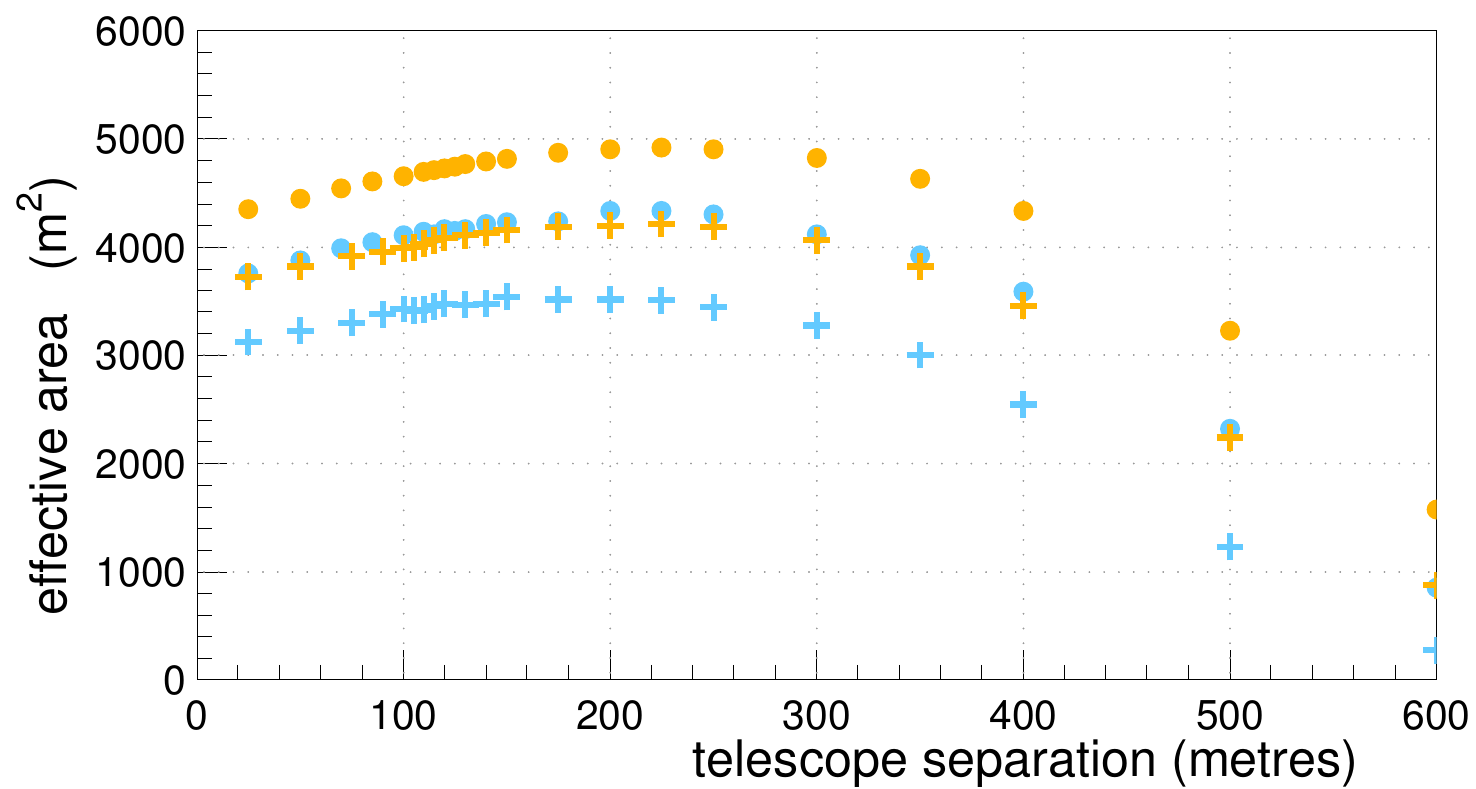}
\end{center}
\caption{\label{fig_telsep}The angular resolution of the four telescope unit (top) and its effective area (bottom) as a function of telescope seperation.}
\end{figure}

The performance of the system for $\gamma$-rays, namely the efficiency of source, core and energy reconstruction is evaluated for different inter-telescope distances between 25 and 600~m. We specially focus on the angular resolution shown here in fig.~\ref{fig_telsep} (top) as a function of the inter-telescope distance. The source position is best reconstructed when the inter-telescope distance lies within an optimum range of around 100-200~m. At higher altitude, this optimum range seems to be slightly narrower and the reconstruction of the source slightly poorer. The same optimum range can be found when looking at shower core reconstruction results (not shown here). The effective area (fig.~\ref{fig_telsep}, bottom) is essentially flat over a large range (100-300~m) of telescope separations with an optimum lying around 200~m. In both plots, one notes that the optimum range is independent of the energy range being considered and has only a slight dependence on the altitude of observation.

 Note that in order to focus on the intrinsic characteristics of the showers and avoid the introduction of too many parameters we have neither added night sky background to the simulations nor performed image cleaning. The impact of the geomagnetic field has been ignored as well. 

\subsection{Possible IACT array configuration}

The results shown in the previous section can then be used to obtain an array configuration. 

\begin{figure}
\begin{center}
\includegraphics [width=0.18\textwidth,angle=270]{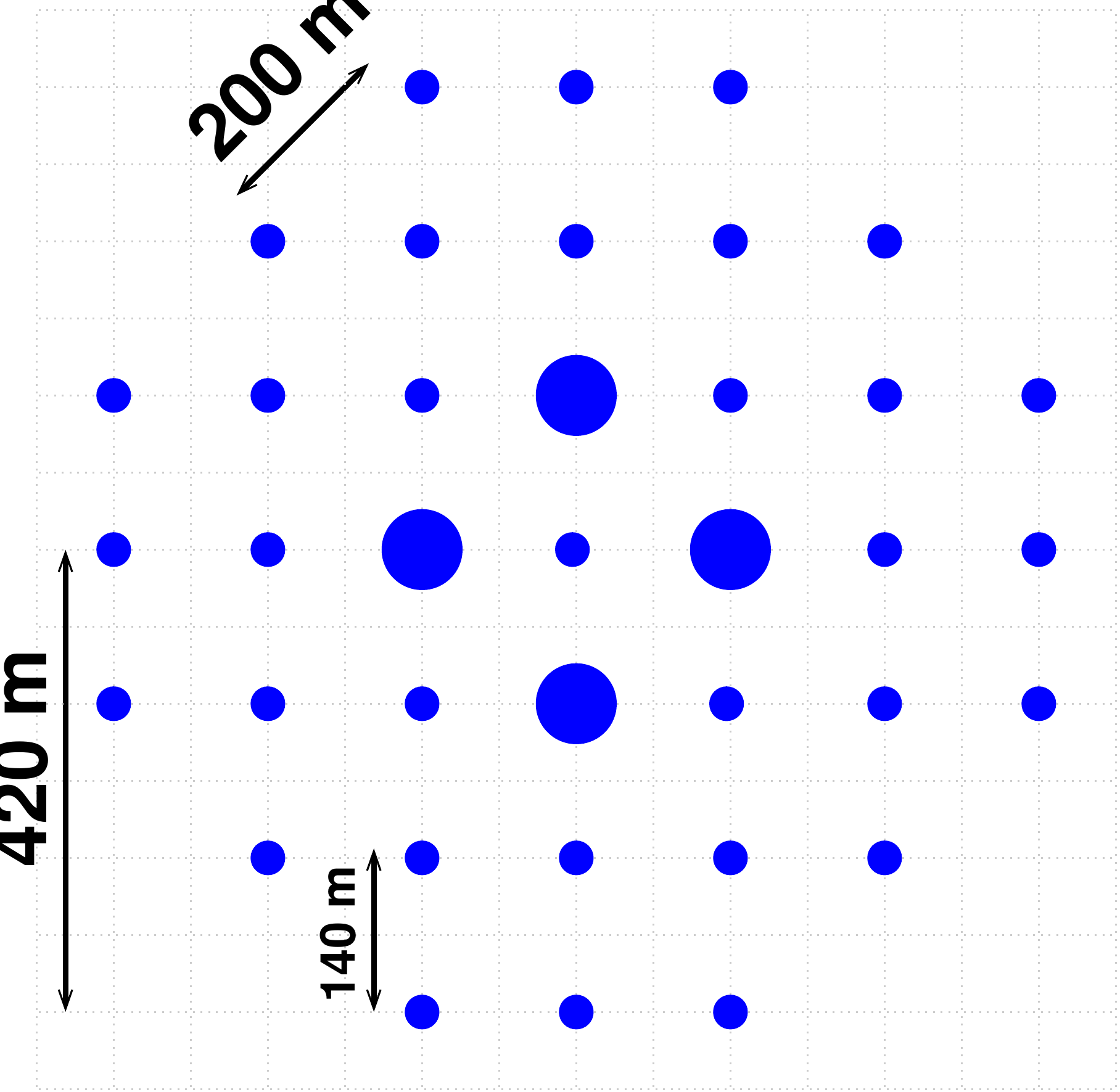}
\includegraphics [width=0.18\textwidth,angle=270]{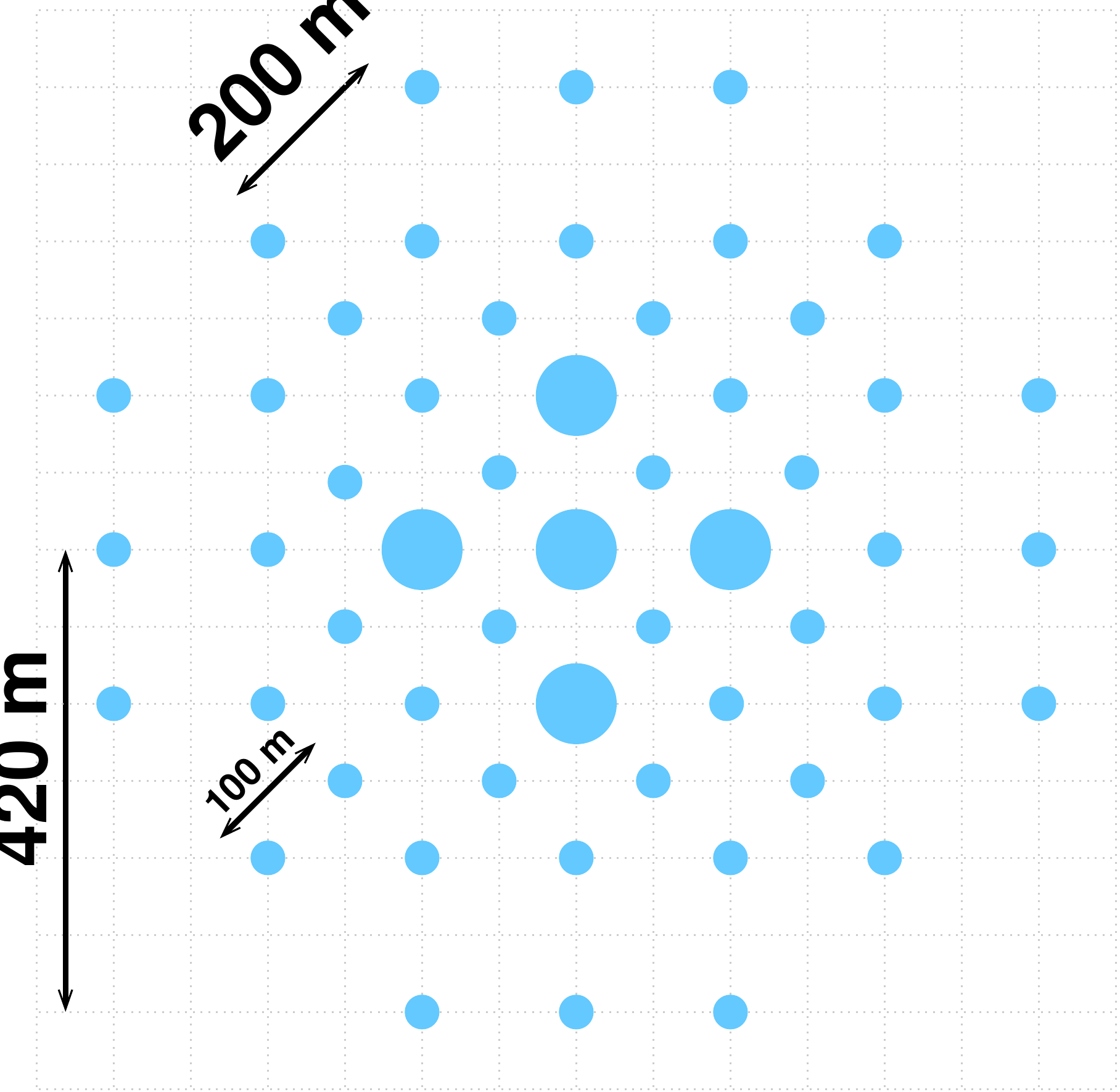}
\caption{\label{fig_configurations} Array configuration 1 (left) and configuration 2 (right) at 1800~m a. s. l..}
\end{center}
\end{figure}

\textbf{Low altitude configurations - }
The two configurations for low altitude observations are shown in fig. \ref{fig_configurations}. In the first configuration, four telescopes of 30~m diameter are placed at the corners of a 200~m sided square for low energy observations. This inter-telescope distance corresponds to the upper edge of the optimum telescope separation range. This choice allows us to optimise parameter reconstruction while keeping the largest possible effective area, with only four telescopes. For the high energy domain a total of 33 medium sized telescopes are spread over a region with a radius of about 400~m. They are distributed in such a way that the inter-telescope distance in the resulting array is of 140~m. This lies in the middle of the optimum telescope separation range. In order to study the effect of a denser array, 16 more medium sized telescopes are added as shown in the right figure and the central telescope is replaced by a large sized telescope. The resulting configuration has an inter-telescope distance of 100~m which corresponds to lower edge of the optimum telescope separation range.

\textbf{High altitude configurations - }
The two configurations are re-scaled\footnote{The system is scaled by a factor corresponding to the ratio of the Cherenkov ring size on the ground at both altitudes. The plots in fig.~\ref{fig_telsep} suggest that there may be ways other than the simple re-scaling of the system to obtain a configuration at high altitude. However, the use of the same configuration allows us to have first comparison between systems at both altitudes.} at higher altitudes so that the inter-telescope distance is 175~m for the large telescopes, 120~m for the medium sized telescopes and 87~m for the denser configuration. 

\section{Performance of possible arrays for $\gamma$-photons}
The performance of the two telescope configurations is studied by uniformly generating $\gamma$-ray showers over a surface of 2400m$\times$2400m at fixed energies and applying the simple trigger described earlier. Figure \ref{fig_configurations} shows the effective area and angular and energy resolutions obtained for the two arrays.
 As can be expected, the effective area of the arrays at high altitude is smaller since the system has been re-scaled and the size of the Cherenkov ring is smaller. This effect is specially visible for higher energies where showers tend to get cut-off by the ground before having fully developed in the atmosphere. 
An angular resolution of around 0.07$^\circ$ is achieved at 1000~GeV. Note that while a four telescope system yields similar angular resolutions for shower falling within a radius of around 150~m, this angular resolution is calculated for all showers generated within a square region of 800m$\times$800m. Similar remarks can be made about the energy resolution of around 7\% achieved at 1~TeV.
One also notes, that the use of a denser array (configuration 2) does not seem to have any impact on the reconstruction of these parameters. Finally, the reconstruction capabilites of the arrays seem to improve slightly at lower altitude. Further details of this study can be found in \cite{phdthesis}.
\begin{figure}
\begin{center}
\includegraphics [width=0.34\textwidth]{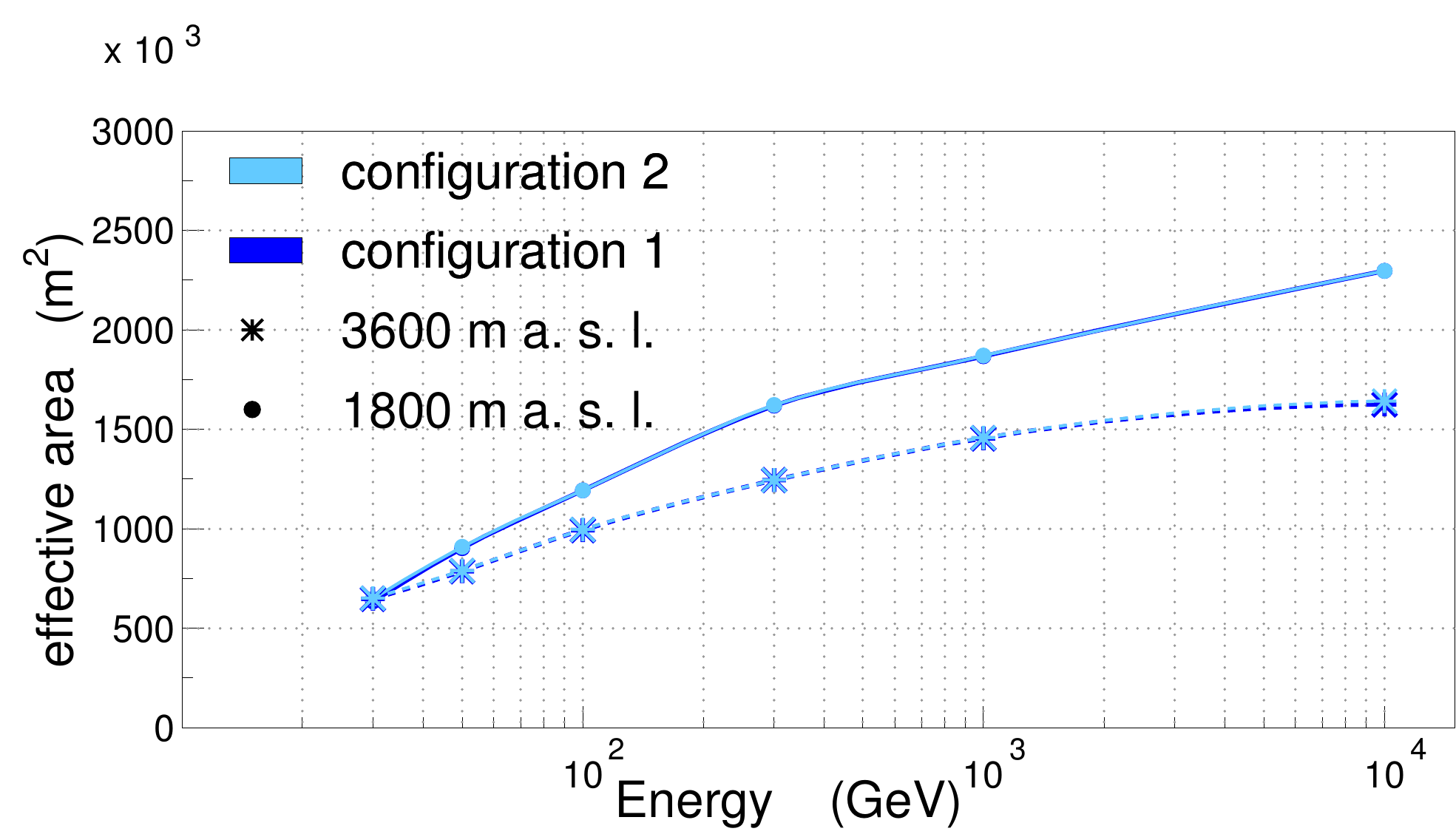}
\includegraphics [width=0.34\textwidth]{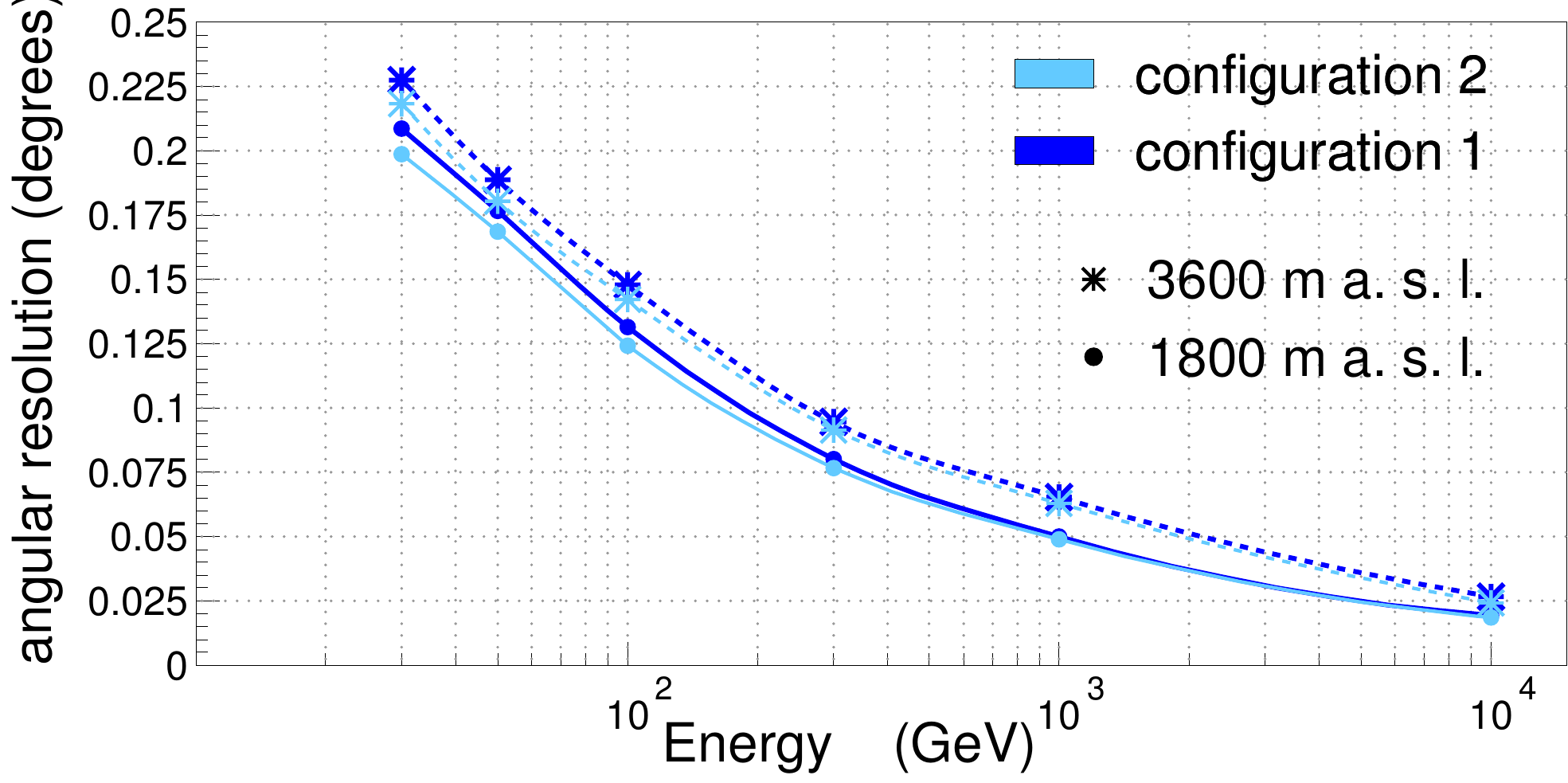}
\includegraphics [width=0.34\textwidth]{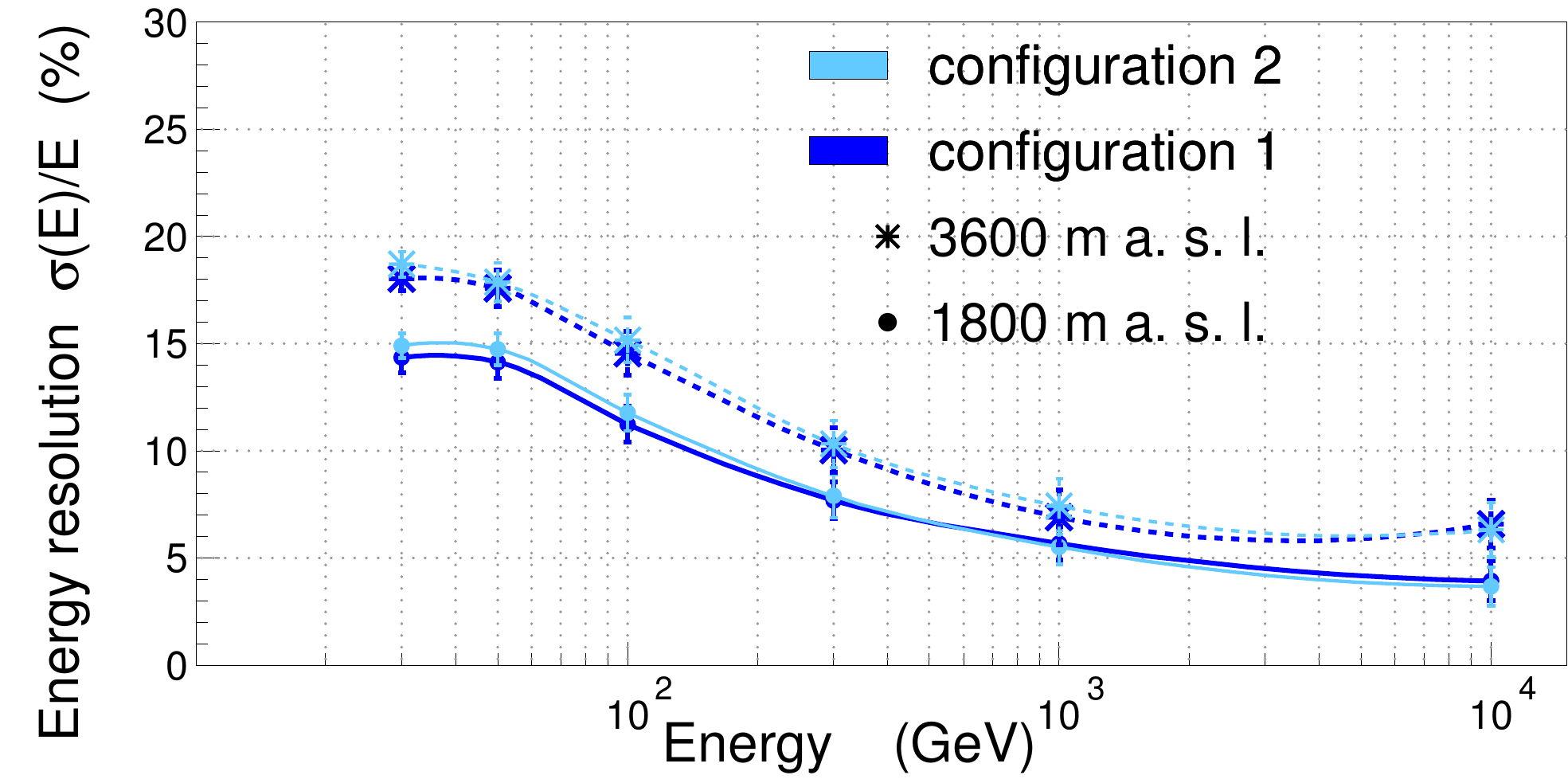}
\caption{\label{fig_results} The effective area and angular and energy resolutions of the two IACT configurations at 1800~m and 3600~m a. s. l..}
\end{center}
\end{figure}

\end{document}